\documentclass[prd,nofootinbib,showpacs,preprint]{revtex4}

\usepackage{graphicx}
\DeclareGraphicsExtensions{.pdf}  % include pdf files by default
\usepackage{color}
\usepackage{epsfig}
\usepackage[autostyle]{csquotes}
\usepackage{multirow}
\usepackage{hyperref}
\usepackage{amsmath}

\newcommand{\be}{\begin{equation}}
\newcommand{\ee}{\end{equation}}
\newcommand{\bea}{\begin{eqnarray}}
\newcommand{\eea}{\end{eqnarray}}
\newcommand{\nn}{\nonumber}

\def\s1{\hat s}

\begin{document}

\title{ Implications of Fermionic Dark Matter on recent neutrino oscillation data{\footnote{I would like to thank Rukmani Mohanta for helpful suggestions. This work is supported by DST-Inspire Fellowship division - IF130927.}}}
\author{Shivaramakrishna Singirala }
\affiliation{\,School of Physics, University of Hyderabad, 
              Hyderabad - 500046, India  }

\begin{abstract}

We investigate flavor phenomenology and dark matter in the context of scotogenic model. In
this model, the neutrino masses are generated through radiative corrections at one-loop level. Considering the neutrino mixing matrix to be of tri-bimaximal form with additional perturbations to accommodate the recently observed non-zero value of reactor mixing angle $\theta_{13}$, we obtain the
relation between various neutrino oscillation parameters and the model parameters. Working in degenerate heavy neutrino mass spectrum, we obtain light neutrino masses obeying normal hierarchy and also study the relic abundance of fermionic dark matter candidate including coannihilation effects. A viable parameter
space is thus obtained, consistent with neutrino oscillation data, relic abundance and various lepton flavor violating decays such as $\ell_\alpha\to\ell_\beta\gamma$ and $\ell_\alpha \to 3 \, \ell_\beta$. 
\end{abstract}
\maketitle

\section{Introduction}
Standard model (SM) of particle physics has been very successful in explaining physics at the fundamental level. However there are still many open questions for which it does not provide any  satisfactory answer. The existence of dark matter (DM) and the observation of non-zero neutrino masses stand as few of the robust evidences for physics beyond the standard model.

Considerable progress has been made in the determination of neutrino mass square differences and mixing parameters from the data of various solar and atmospheric neutrino oscillation experiments. Theoretically, the smallness of neutrino mass can be generally explained by the well known seesaw mechanisms namely: type-I \cite{t1}, type-II \cite{t2}, type-III \cite{t3} and radiative seesaw \cite{ma}. 
%The neutrino mixing is described by unitary Pontecorvo-Maki-Nakagawa-Sakata (PMNS) matrix $V_{PMNS}$ \cite{pmns}, which can be parameterized in terms of three rotation angles $\theta_{12}^{}$, $\theta_{23}^{}$, $\theta_{13}^{}$ and three CP-violating phases, one Dirac type ($\delta_{CP}$) and two Majorana types ($\rho, \sigma$) as
In standard parametrization, the mechanism of mixing can be described by unitary Pontecorvo-Maki-Nakagawa-Sakata (PMNS) matrix $V_{PMNS}$ \cite{pmns} written in terms of three rotation angles $\theta_{12}^{}$, $\theta_{23}^{}$, $\theta_{13}^{}$ and  three CP-violating phases namely $\delta_{CP}$ (Dirac type) and $\rho, \sigma$ (Majorana type) as
\be
V_{PMNS} \equiv  U_{PMNS} \cdot P_\nu = \left( \begin{array}{ccc} c^{}_{12} c^{}_{13} & s^{}_{12}
c^{}_{13} & s^{}_{13} e^{-i\delta_{CP}} \\ -s^{}_{12} c^{}_{23} -
c^{}_{12} s^{}_{13} s^{}_{23} e^{i\delta_{CP}} & c^{}_{12} c^{}_{23} -
s^{}_{12} s^{}_{13} s^{}_{23} e^{i\delta_{CP}} & c^{}_{13} s^{}_{23} \\
s^{}_{12} s^{}_{23} - c^{}_{12} s^{}_{13} c^{}_{23} e^{i\delta_{CP}} &
-c^{}_{12} s^{}_{23} - s^{}_{12} s^{}_{13} c^{}_{23} e^{i\delta_{CP}} &
c^{}_{13} c^{}_{23} \end{array} \right) P^{}_\nu \;,\label{standpara}
\ee
where $c^{}_{ij}\equiv \cos \theta^{}_{ij}$, $s^{}_{ij} \equiv \sin
\theta^{}_{ij}$ and $P_\nu^{} \equiv \{ e^{i\rho}, e^{i\sigma}, 1\}$ is
a diagonal phase matrix. The mixing angles as well as the mass square differences have been well constrained by various neutrino oscillation experiments. 
Recently Daya Bay \cite{daya-bay1, daya-bay2}, RENO \cite{reno} and T2K \cite{t2k} collaborations have precisely measured the reactor mixing angle $\theta_{13}$  with a moderately large value.
However, there are several missing pieces such as the neutrino mass hierarchy, the magnitude of the CP violating phase $\delta _{CP}$, the absolute scale of the neutrino mass, and the nature of neutrinos (whether Dirac or Majorana). Various neutrino oscillation parameters derived from global analysis of recent oscillation data  taken from Ref. \cite{osc}  are presented in Table-I.
\begin{table}[htb]
\begin{center}
\vspace*{0.1 true in}
\begin{tabular}{|c|c|c|}
\hline
 Mixing Parameters & Best Fit value & $ 3 \sigma $ Range  \\
\hline
$\sin^2 \theta_{12} $ &~ $0.323$ ~& ~$ 0.278 \to 0.375 $~\\

$\sin^2 \theta_{23}  $ (NO) &~ $0.567$ ~& ~$ 0.392 \to 0.643 $~\\

$\sin^2 \theta_{13} $ (NO) &~ $0.0234$ ~& ~$ 0.0177 \to 0.0294 $~\\

$\delta_{\rm CP}$ (NO) & ~$1.34 \pi$ & $ ~(0 \to 2 \pi)~ $\\

$\Delta m_{31}^2/ 10^{-3}~ {\rm eV}^2 ~({\rm NO}) $~ &~ $ 2.48
$ & $ 2.3 \to 2.65 $ \\

$\Delta m_{21}^2/ 10^{-5} ~{\rm eV}^2 $ & $ 7.60 $ & $ 7.11 \to 8.18 $ \\
\hline
\end{tabular}
\end{center}
\caption{Best-fit values with their  $3\sigma$ ranges of the neutrino oscillation parameters from \cite{osc} where NO indicates normal ordering. }
\end{table}

On the other hand, the particle nature of dark matter is still a mystery till date.  Recent survey of PLANCK \cite{planck} reveal that DM constitutes about 26.8\% of the total energy budget of the Universe. Various cosmological observations suggest that this unknown particle is non-relativistic in nature and is stable on cosmological time scales. Numerous beyond SM scenarios study DM phenomenology by imposing additional discrete symmetry such as R-parity, $Z_2$ symmetry etc. Weakly Interacting Massive Particle (WIMP miracle) is the best motivated candidate of DM.
They are massive particles with cross section approximately the order of weak interaction cross section.

It would be interesting to study the extensions of standard model that can relate these two issues. Scotogenic model proposed by Ma \cite{ma} is one among such frameworks in which neutrino mass generation involves the interaction with dark matter. In this model an unbroken discrete symmetry forbids neutrino attaining a tree level mass and also assures the stability of DM particle. It is a suitable platform to simultaneously explain neutrino oscillation data and DM phenomenology. 

In this work, we consider the scotogenic model to correlate some of the neutrino oscillation parameters, like the mass square differences and the mixing angles with the model parameters. We examine the neutrino radiative mass matrix using the mixing matrix of TBM type with added perturbation to achieve large $\theta_{13}$. We solve for suitable flavor structure to study neutrino phenomenology. We then use the best fit values on neutrino oscillation parameters to constrain the parameter space of this model. In addition, we study DM relic abundance choosing the lightest odd particle as the DM candidate. We scan over entire parameter space of the model imposing the constraints from neutrino data, DM observables and lepton flavor violating decays.

The paper is organized as follows. In section II we describe the scotogenic model. In section III we diagonalize the neutrino radiative mass matrix and obtain solutions to explain neutrino oscillation data. The fermionic DM relic abundance considering the coannihilation effects is studied in section IV and then in section V we estimate the branching ratios of various LFV decays. We conclude our discussion in section VI.

\section{Scotogenic model}
Scotogenic model is a minimal extension of standard model with an additional inert scalar doublet $\eta$ and three heavy Majorana right-handed neutrinos $N_i$ ($i=1,2,3$). The potential is imposed with a discrete symmetry under which all the new particles i.e., $N_i$ and $\eta$ are odd and SM particles are even. The unbroken discrete  symmetry guarantees the coupling of the inert doublet to  fermions vanish and doesn't get a vacuum expectation value (VEV). While the SM Higgs doublet $\phi$ obtains a VEV $\langle \phi^0 \rangle = v$ by the spontaneous symmetry breaking of $SU(2)_L \times U(1)_Y$ global symmetry. This model is rich in phenomenology providing scalar and fermionic dark matter candidates. Scalar dark matter in this model has been studied extensively in literature \cite{lopez,barbieri,Gustafsson}.

The scalar potential of this model is given by \cite{suematsu}
\begin{eqnarray}
V &=& m_{\phi}^{2} \phi^\dagger \phi + m_{\eta}^{2} \eta^\dagger \eta + {1 \over 2} 
\lambda_1 (\phi^\dagger \phi)^{2} + {1 \over 2} \lambda_2 
(\eta^\dagger \eta)^{2} + \lambda_3 (\phi^\dagger \phi)(\eta^\dagger \eta) 
\nonumber \\ &+& \lambda_4 (\phi^\dagger \eta)(\eta^\dagger \phi) + 
{1 \over 2} \lambda_5 [(\phi^\dagger \eta)^{2} +(\eta^\dagger \phi)^{2}],
\end{eqnarray} 
where the two scalar doublets $\phi$ and $\eta$ are defined as

\be
\phi =    \left ( \begin{array}{c}
 \phi^+     \\
\phi^0    \\
\end{array}
\right ),
\qquad
\eta =    \left ( \begin{array}{c}
 \eta^+     \\
\eta^0    \\
\end{array}
\right ).
\ee
After spontaneous symmetry breaking, the masses of the charged component ($\eta^+$) and neutral components of $\eta^0 = (\eta_R + i \eta_I)/\sqrt{2}$ are given by
\begin{eqnarray}
&&m_{\eta^+}^{2}   = m_{\eta}^{2}+\lambda_{3}v^{2} ,\nonumber \\
&&m_{R}^{2}    = m_{\eta}^{2}+(\lambda_{3}+\lambda_{4}+\lambda_{5})v^{2}, 
\nonumber \\
&&m_{I}^{2}    = m_{\eta}^{2}+(\lambda_{3}+\lambda_{4}-\lambda_{5})v^{2}.
\end{eqnarray}
The Yukawa Lagrangian of this model is \cite{suematsu}
\begin{equation}
\mathcal{L}_N=\overline{N_i}i\partial\!\!\!/\!\:P_RN_i
+\left(D_\mu\eta\right)^\dag\left(D^\mu\eta\right)
-\frac{M_i}{2}\overline{N_i\:\!^c}P_RN_i+h_{\alpha
 i}\overline{\ell_\alpha}\eta^\dag P_RN_i+\mathrm{h.c.},
\label{yuk}
\end{equation}
where $h_{\alpha i}$ are the Yukawa couplings, $\alpha$ denotes the lepton flavor and $M_i$ are the masses of heavy neutrinos $N_i$.

In this model, neutrinos get their mass by loop correction called \enquote{radiative seesaw mechanism}. The corresponding neutrino mass matrix is given by 
\be
({\cal M}_\nu)_{\alpha\beta}=\sum_{i=1}^3h_{\alpha i}h_{\beta i}\Lambda_i,
\label{lpmass1}
\ee
where $\Lambda_i$ is defined as
\be
\Lambda_i=\frac{\lambda_5 v^2}{8\pi^2M_i}I\left(r_i\right), 
\qquad
I(x)=\frac{x^2}{1-x^2}\left(1+\frac{x^2}{1-x^2}\ln x^2\right),
\label{lpmass2}
\ee
Here the parameters $r_i$ are defined as $r_i = M_i/m_0$ and ${m_0}^2 = ({m_{R}}^2 + {m_I}^2)/2$. We take $ \lambda_{5} \sim  10^{-10}$, a very small value in order to have correct neutrino masses and also probe for lepton flavor violation \cite{schmidt, suematsu, vicente1,vicente2}. We now diagonalize the radiative mass matrix (\ref{lpmass1}) using PMNS matrix to explain neutrino oscillation data.
%%%%%%%%%%%%%%%%%%%%%%%%%%%%%%%%%%%%%%%%%%%%%%%%%%%%%%%%%%%%%%%%%%%%%%%%%%%%%
\section{Neutrino phenemenology}
Various neutrino experiments confirmed that neutrinos have tiny mass and they oscillate from one flavor to another as they propagate. The phenomenon of neutrino oscillation is described by solar ($\theta_{12}$), atmospheric ($\theta_{23}$) and reactor ($\theta_{13}$) mixing angles. Of these three rotation angles, two are large ($\theta_{12}$ and $\theta_{23}$), and one is not so large ($\theta_{13}$). Originally, it was believed that the reactor mixing angle would be very small and with this motivation numerous models were proposed which are generally based on some discrete flavor symmetries such as $S_3$, $S_4$, $A_4$, etc \cite{sym} to explain the neutrino mixing pattern. For instance, the tri-bimaximal (TBM) mixing pattern \cite{tbm}, a well motivated model having $\sin^2 \theta_{12} = \frac{1}{3}$ and $\sin^2 \theta_{23} = \frac{1}{2}$ which can be expressed in a generalized form as
\begin{equation}
U_\nu^0=\left(\begin{array}{ccc}
\cos\theta & \sin\theta & 0 \\
-\frac{\sin\theta}{\sqrt 2} & \frac{\cos\theta}{\sqrt 2} & 
\frac{1}{\sqrt 2}\\
\frac{\sin\theta}{\sqrt 2} & -\frac{\cos\theta}{\sqrt 2} & 
\frac{1}{\sqrt 2}\\
\end{array}\right),
\label{gen}
\end{equation}
with $\theta \simeq 35^{\circ}$. However, in TBM mixing pattern the value of $\theta_{13}$ turn out to be zero. After the experimental evidence of moderately large $\theta_{13}$, it was found that adding suitable perturbation terms to the TBM mixing pattern can still 
describe the neutrino mixing pattern with sizeable $\theta_{13}$. As discussed in \cite{sruthilaya}, here we consider a simple perturbation matrix, i.e., a rotation matrix in 13 plane, which can provide the required corrections to the various mixing angles of TBM
mixing matrix. Assuming the charged lepton mass matrix is diagonal (i.e., identity matrix), one can write the PMNS mixing matrix, which relates the flavor eigenstates to the corresponding mass eigenstates as
\be
U_{PMNS} = U_\nu^0   \left ( \begin{array}{ccc}
 \cos \varphi    & 0 & e^{-i  \zeta} \sin \varphi \\
0    & 1    & 0 \\
 -e^{i \zeta} \sin \varphi  & 0   & \cos \varphi\\
\end{array}
\right ).
\label{per}
\ee
In our work, we consider the phase $\zeta$ to be zero for convenience. Now we diagonalize the mass matrix (\ref{lpmass1}) by the mixing matrix (\ref{per}) using the relation
$
U_{PMNS}^T{\cal M}_\nu U_{PMNS  }={\rm diag}(m_1,m_2,m_3).
$
This in turn provides the following conditions (vanishing off-diagonal elements of the mass
matrix) to be satisfied:
\begin{subequations}
\bea
&&\sum_{i=1}^3\frac{h_{ei}^2}{2}\sin 2\theta\cos \varphi +\frac{h_{ei}(h_{\mu i}-h_{\tau i})}{\sqrt2}
\cos 2\theta \cos \varphi-\frac{(h_{\mu i}-h_{\tau i})^2}{4}\sin 2\theta \cos \varphi\nn \\
&& - \frac{h_{ei}(h_{\mu i}+h_{\tau i})}{\sqrt2}\sin\theta \sin \varphi-\frac{(h_{\mu i}^2-h_{\tau i}^2)}{2}\cos\theta\sin\varphi=0,  \label{diag-a}\\
&&\sum_{i=1}^3\frac{h_{ei}^2}{2}\sin 2\theta\sin \varphi +\frac{h_{ei}(h_{\mu i}-h_{\tau i})}{\sqrt2}
\cos 2\theta \sin \varphi-\frac{(h_{\mu i}-h_{\tau i})^2}{4}\sin 2\theta \sin \varphi \nn \\
&& + \frac{h_{ei}(h_{\mu i}+h_{\tau i})}{\sqrt2}\sin\theta \cos \varphi+\frac{(h_{\mu i}^2-h_{\tau i}^2)}{2}\cos\theta\cos\varphi=0, \label{diag-b} \\
&&\sum_{i=1}^3\frac{h_{ei}^2}{2}\cos^2\theta \sin 2\varphi -\frac{h_{ei}(h_{\mu i}-h_{\tau i})}{2\sqrt2}
\sin 2\theta \sin 2\varphi+\frac{(h_{\mu i}-h_{\tau i})^2}{4}\sin^2\theta \sin 2\varphi \nn \\
&& + \frac{h_{ei}(h_{\mu i}+h_{\tau i})}{\sqrt2}\cos\theta \cos 2\varphi-\frac{(h_{\mu i}^2-h_{\tau i}^2)}{2}\sin\theta\cos2\varphi
-\frac{(h_{\mu i}+h_{\tau i})^2}{4}\sin2\varphi=0. ~~
\label{diag-c}
\eea 
\end{subequations}
%%%%%%%%%%%%%%%%%%%%%%%%%%%%%%%%%%%%%%%%%%%%%%%%%%%%%%%%%%%%%%%%%%%%%%%%%%%%%
The neutrino mass eigenvalues are given by
\bea
m_1&=&\sum_{i=1}^3 (h_{ei}^2\cos^2\theta \cos^2\varphi-\frac{1}{\sqrt 2} 
h_{ei}(h_{\mu i}-h_{\tau i})\sin 2\theta \cos^2\varphi-\frac{1}{\sqrt 2} 
h_{ei}(h_{\mu i}+h_{\tau i})\cos \theta \sin 2\varphi\nonumber \\
&+&\frac{1}{2}(h_{\mu i}+h_{\tau i})^2\sin^2\varphi+\frac{1}{2}(h_{\mu i}^2-h_{\tau i}^2)\sin\theta \sin2\varphi+\frac{1}{2}(h_{\mu i}-h_{\tau i})^2\sin^2\theta\cos^2\varphi )\Lambda_i \;,\nonumber \\
m_2&=&\sum_{i=1}^3(h_{ei}^2\sin^2\theta+\frac{1}{\sqrt2}h_{ei}(h_{\mu i}-h_{\tau i})\sin2\theta+\frac{1}{2}(h_{\mu i}-h_{\tau i})^2\cos^2\theta )\Lambda_i\;,\nonumber  \\
m_3 & = &\sum_{i=1}^3(h_{ei}^2\cos^2\theta \sin^2\varphi-\frac{1}{\sqrt 2} 
h_{ei}(h_{\mu i}-h_{\tau i})\sin 2\theta \sin^2\varphi+\frac{1}{\sqrt 2} 
h_{ei}(h_{\mu i}+h_{\tau i})\cos \theta \sin 2\varphi\nonumber \\
& +&\frac{1}{2}(h_{\mu i}+h_{\tau i})^2\cos^2\varphi-\frac{1}{2}(h_{\mu i}^2-h_{\tau i}^2)\sin\theta \sin2\varphi+\frac{1}{2}(h_{\mu i}-h_{\tau i})^2\sin^2\theta\sin^2\varphi)\Lambda_i\;.
\label{masseigen}
\eea 
%%%%%%%%%%%%%%%%%%%%%%%%%%%%%%%%%%%%%%%%%%%%%%%%%%%%%%%%%%%%%%%%%%%%%%%%%%%%%
Solving (\ref{diag-a}), (\ref{diag-b}) and substituting in (\ref{diag-c}), we obtain two solutions given by
\bea \label{soln}
1.\hspace{0.8 cm}&& h_{\mu i_1}\not= -h_{\tau i_1},\quad\tan{\theta}=\frac{(h_{\tau i_1}-h_{\mu i_1})}{\sqrt2 
h_{e i_1}}, \nonumber\\
&&\tan2\varphi=\frac{-\left(\frac{h_{e i_1}(h_{\mu i_1}+h_{\tau i_1})}{\sqrt2}\cos\theta-\frac{(h_{\mu i_1}^2-h_{\tau i_1}^2)}{2}\sin\theta \right)}{\left(\frac{h_{e i_1}^2}{2}\cos^2\theta-\frac{h_{e i_1}(h_{\mu i_1}-h_{\tau i_1})}{2\sqrt2}\sin2\theta+\frac{(h_{\mu i_1}-h_{\tau i_1})}{4}^2\sin^2\theta-\frac{(h_{\mu i_1}+h_{\tau i_1})^2}{4} \right)},  \nonumber 
\eea
\bea
2.\hspace{1cm}  h_{\mu i_2}=-h_{\tau i_2},\quad\tan{\theta}=\frac{h_{e i_2}}{\sqrt2 h_{\mu i_2}}, \hspace{7.5cm}
\eea 
%%%%%%%%%%%%%%%%%%%%%%%%%%%%%%%%%%%%%%%%%%%%%%%%%%%%%%%%%%%%%%%%%%%%%%%%%%%%
where $i_1, i_2$ can take any value of $i (=1,2,3)$. As shown in Ref. \cite{sruthilaya}, the above mixing matrix can explain recent neutrino oscillation data with the unperturbed mixing as TBM type (i.e., with $\theta = 35^\circ$)  and the perturbed angle $\varphi = {12}^{\circ}$, which accommodates the experimentally measured mixing angles.
Thus,  eqn. (\ref{soln}) further gets simplified to three simple solutions and the obtained flavor structure written in terms of $h_{e i} (=h_i)$ in a matrix labelled with the lepton flavor $\alpha$ as row index and $i=1,2,3$ denote the column index given by
\be
h_{\alpha i} =    \left ( \begin{array}{ccc}
 h_1    &  h_2 & h_3 \\
-0.68 ~h_1    & h_2    &  3.56 ~h_3\\
 0.31 ~h_1  &  -h_2  & 4.55 ~h_3\\
\end{array}
\right ).
\label{flav}
\ee
Here $i_1 = 1,3$ and $i_2 = 2$ is assumed so that the mass eigenvalues (\ref{masseigen}) get non-zero contribution given as
\begin{eqnarray}
&&m_{1} = c_1  (h_1^2 \Lambda_1) ,\nonumber\\
&&m_{2} = c_2 (h_2^2 \Lambda_2),\nonumber\\
&&m_{3} = c_3  (h_3^2 \Lambda_3), 
\label{con}
\end{eqnarray}
where the coefficients $c_1 = 1.55$, $c_2 = 3.04$, $c_3 = 34.44$. Thus, the flavor structure (\ref{flav}) is suitable to explain normal hierarchy i.e., $(m_3 \gg m_2 > m_1)$ provided an assumption that $N_1$ and $N_2$ are degenerate. Imposing the best fit values given in Table-1, the constraint from neutrino mass square differences are given by
\begin{eqnarray}
&&\left[(c_2 h^2_{2})^2 - (c_1 h^2_{1})^2\right]\Lambda_1^2 = 7.6 \times 10^{-5} ~{\rm eV^2}, \nonumber\\
&&\left[(c_3 h^2_{3} \Lambda_3)^2 - (c_2 h^2_{2} \Lambda_1)^2\right] = 2.4 \times 10^{-3} ~{\rm eV^2}.
\label{sq}
\end{eqnarray}
Thus, we have a free parameter space spanned by $h_i, r_{1,3}$  and $M_{1,3} $. We now proceed to constrain the parameter space with the DM relic abundance choosing the lightest one of the odd particles as a DM candidate.
%%%%%%%%%%%%%%%%%%%%%%%%%%%%%%%%%%%%%%%%%%%%%%%%%%%%%%%%
\section{Relic abundance}
We choose $N_1$ as the lightest odd particle and since $N_2$
is its degenerate partner, the relic abundance gets contributions from annihilation as well
as coannihilation channels. To include the coannihilation effects, we adopt the procedure given in \cite{griest} in the estimation of relic abundance. We introduce a parameter $\delta$ given by $\delta\equiv (M_2-M_1)/M_1$ which depicts the mass splitting ratio of the degenerate neutrinos.  
The effective cross section $\sigma_{\rm eff}$ including contributions from coannihilations is given by
\begin{eqnarray}
\sigma_{\rm eff}&=& 
\frac{g_{N_1}^2}{g_{\rm eff}^2}\sigma_{N_1N_1}+
2\frac{g_{N_1}g_{N_2}}{g_{\rm eff}^2}
\sigma_{N_1N_2}(1+\delta)^{3/2}e^{-\delta x}
+\frac{g_{N_2}^2}{g_{\rm eff}^2}\sigma_{N_2N_2}
(1+\delta)^3e^{-2\delta x}, \nonumber \\
g_{\rm eff}&=&g_{N_1}+g_{N_2}(1+\delta)^{3/2}e^{-\delta x}.
\label{sig}
\end{eqnarray}
Here $g_{\rm eff}$ denotes the effective degrees of freedom, $g_{N_{1,2}}$ are the number of degrees of freedom for Majorana fermion and $x = {M_1}/T$, where $T$ is the temperature. The (co)annihilation cross section of $N_{i}$ and $N_{j}$ is given by \cite{suematsu}
%%%%%%%%%%%%%%%%%%%%%%%%%%%%%%%%%%%%%%%%%%%%%%%%%%%%%%%%%%%%%%%%%%%%%%%%%%%%%
\begin{eqnarray}
\sigma_{N_{i}N_{j}}|v_{\rm rel}|&=&\frac{1}{8\pi}
\frac{M_1^2}{(M_1^2+ m_0^2)^2}
\left[1 + \frac{m_0^4- 3m_0^2 M_1^2 -M_1^4}{3(M_1^2 
+m_0^2)^2}v_{\rm rel}^2 \right]
\times 
\sum_{\alpha,\beta}(h_{\alpha i}h_{\beta j}
-h_{\alpha j}h_{\beta i})^2 \nonumber \\
&+&\frac{1}{12 \pi}\frac{M_1^2(M_1^4+m_0^4)}{(M_1^2+m_0^2)^4}v_{\rm rel}^2
\sum_{\alpha,\beta}h_{\alpha i}h_{\alpha j}h_{\beta i}h_{\beta j}.
\end{eqnarray}
%%%%%%%%%%%%%%%%%%%%%%%%%%%%%%%%%%%%%%%%%%%%%%%%%%%%%%%%%%%%%%%%%%%%%%%%%%%
In the above expression $i, j$ can be 1 or 2 and $v_{\rm rel}$ represents the relative velocity of annihilating particles. The effective annihilation cross section is defined as $\sigma_{\rm eff}|v_{\rm rel}|=a_{\rm eff}+b_{\rm eff}v_{\rm rel}^2$. The coefficients $a_{\rm eff}$ and $b_{\rm eff}$ for the obtained flavor structure (\ref{flav}) are given by
\begin{equation}
a_{\rm eff}=\frac{1}{16\pi}
\frac{M_1^2}{(M_1^2+ m_0^2)^2}(s_{12} h^2_{1} h^2_{2}), 
\end{equation} 
\begin{eqnarray}
b_{\rm eff} &=& \frac{1}{48 \pi}\frac{M_1^2(M_1^4+m_0^4)}{(M_1^2+m_0^2)^4}\left[(s_{1} h^4_{1} +  s_2 h^4_{2})\right] \nonumber\\
&+& \frac{1}{16\pi}
\frac{M_1^2}{(M_1^2+ m_0^2)^2}\left[\frac{m_0^4- 3m_0^2 M_1^2 -M_1^4}{3(M_1^2 
+m_0^2)^2}\right](s_{12} ~ h^2_{1} h^2_{2}),
\end{eqnarray}
where $s_1 = 2.42, s_2 = 9.24$ and $s_{12} = 9.47$. Now the thermally averaged cross section is
given as $\langle\sigma_{\rm eff}|v_{\rm rel}|\rangle=a_{\rm eff}+ 6 b_{\rm eff}/x$. If the decoupling temperature is given by $T_f = {M_1}/{x_f}$,
the relic abundance can be estimated by 
%%%%%%%%%%%%%%%%%%%%%%%%%%%%%%%%%%%%%%%%%%%%%%%%%%%%%%%%%%%%%%%%%%%%%%%%%%%%%
\begin{equation}
\Omega_{N_1} h^2 = \frac{1.07 \times 10^9 \rm GeV^{-1}}{g_\star ^{1/2} m_{\rm pl}} \frac{1}{J(x_f)}\,,
\end{equation}
where $m_{\rm pl}=1.22\times 10^{19}$~GeV and $g_\star = 106.75$ and $J(x_f)$ is given by
\begin{equation}
J(x_f)= \int_{x_f}^ \infty \frac{\langle \sigma_{\rm eff} |v_{\rm rel}| \rangle _{\rm eff}}{x^2} \hspace{.2cm} dx. 
\end{equation}  
%%%%%%%%%%%%%%%%%%%%%%%%%%%%%%%%%%%%%%%%%%%%%%%%%%%%%%%%%
Using the first relation in eqn. (\ref{sq}), we eliminate $h_2$ and since $N_1$ is the lightest odd particle, we take $r_1 < 1$ \cite{suematsu, vicente2} and  $|h_{i}| < 1.5$ \cite{schmidt}. Fig. 1  depicts the allowed parameter space $(h_1 , r_1)$ consistent with current bound on relic abundance \cite{planck}. 
%From the figure it is clear that a large range of values for $r_1$ is allowed for $|h_{1}| < 1$. 
Fig. 2 displays the relic abundance as a function of DM mass for various values of $h_1$ at two representative values of $r_1$, i.e., $r_1 = 0.5$ in the left panel and $r_1 = 0.6$ in the right panel. This shows that the mass range of DM mass consistent with current relic abundance is proportional with the parameter $r_1$ and the Yukawa coupling $h_1$.
%%%%%%%%%%%%%%%%%%%%%%%%%%%%%%%%%%%%%%%%%%%%%%%%%%%%%%%%%%%%%%%%%%%%%%%%%%%%
\begin{figure}[htb]
\begin{center}
\includegraphics[width=7.0 cm,height= 5.0 cm, clip]{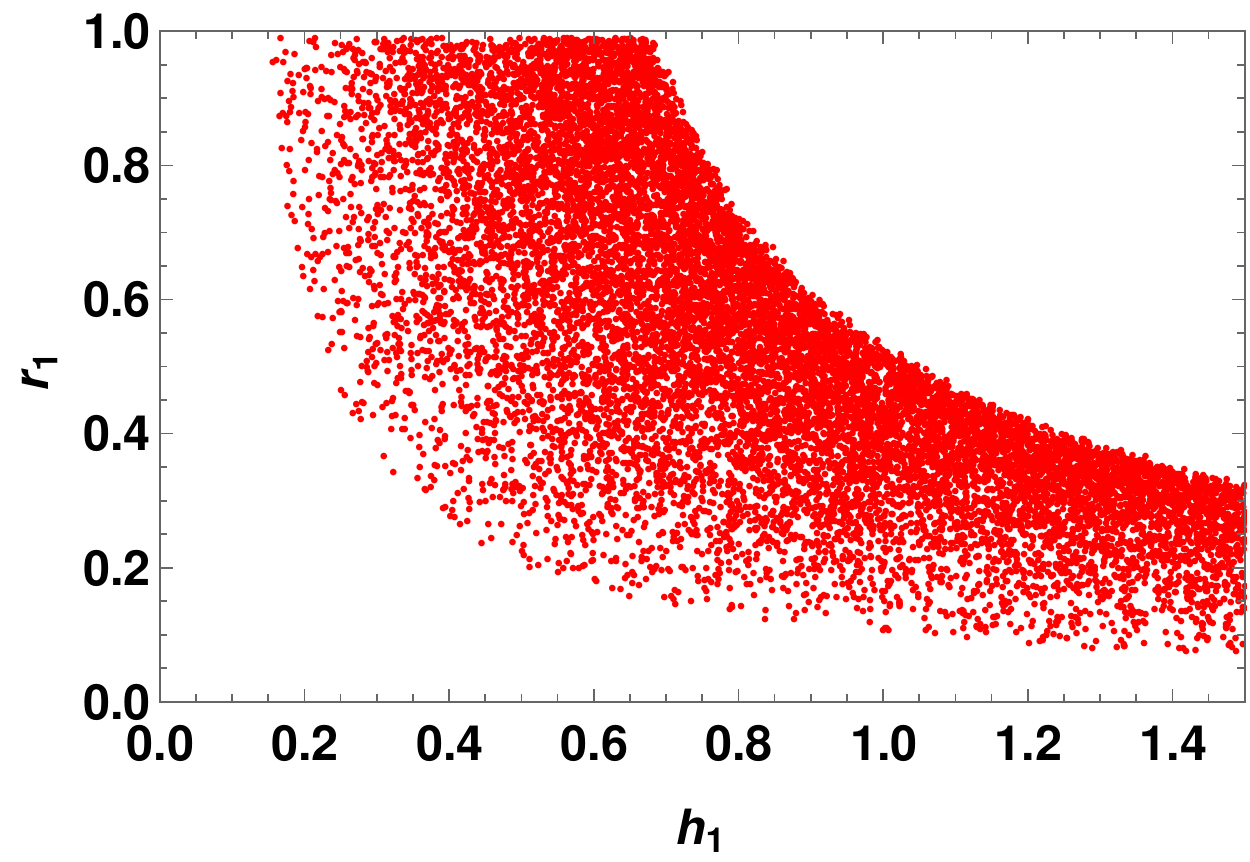}
\caption{Parameter space of $h_{1}$ and $r_1$ consistent with $3\sigma$ relic abundance.}
\end{center}
\label{h1vsx1}
\end{figure}
%%%%%%%%%%%%%%%%%%%%%%%%%%%%%%%%%%%%%%%%%%%%%%%%%%%%%%%%%%%%%%%%%%%%%%%%%%%%%
%%%%%%%%%%%%%%%%%%%%%%%%%%%%%%%%%%%%%%%%%%%%%%%%%%%%%%%%%%%%%%%%%%%%%%%%%%%%
\begin{figure}[htb]
\begin{center}
\includegraphics[width=7.0 cm,height= 5.0 cm, clip]{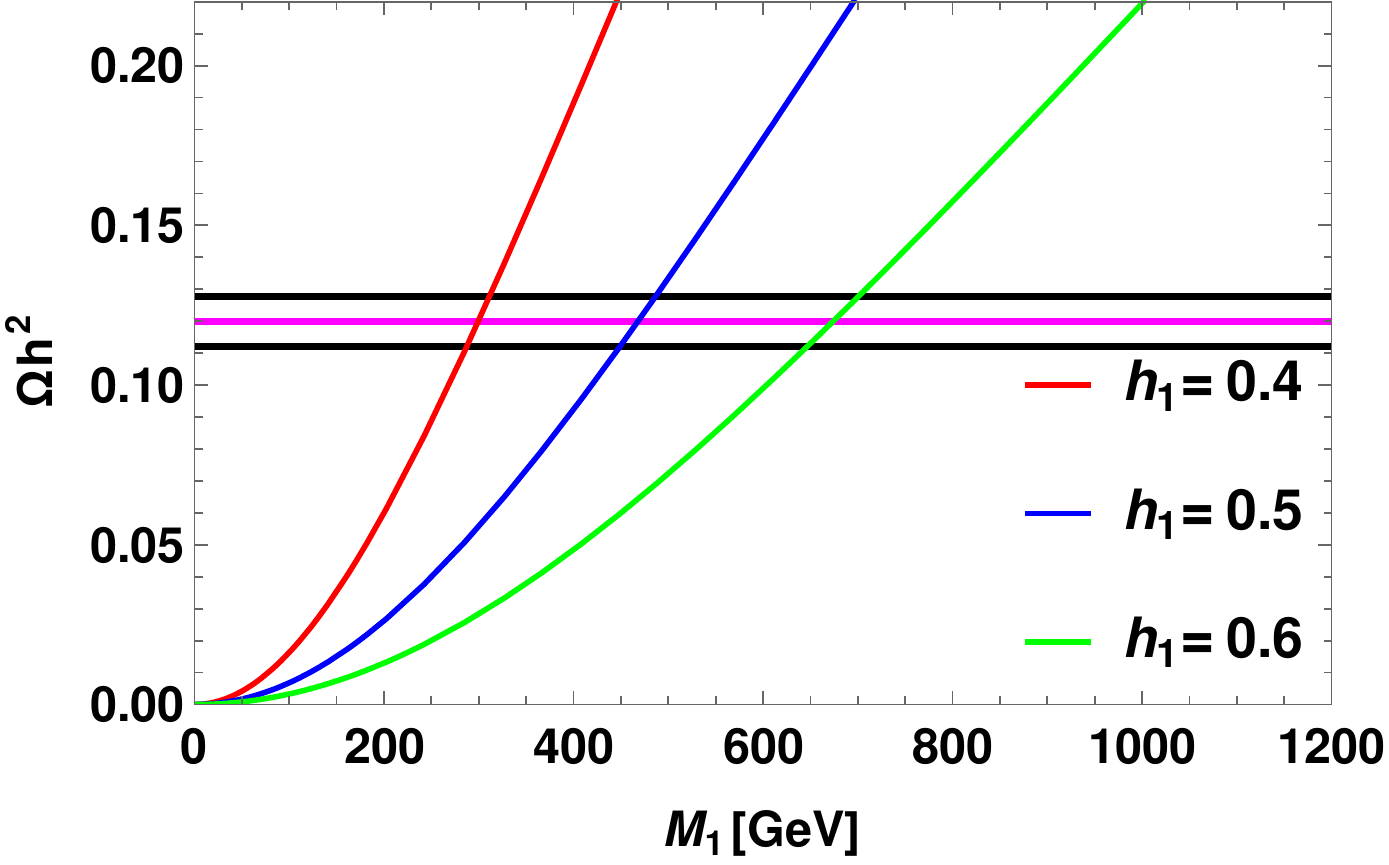}
\quad
\includegraphics[width=7.0 cm,height= 5.0 cm, clip]{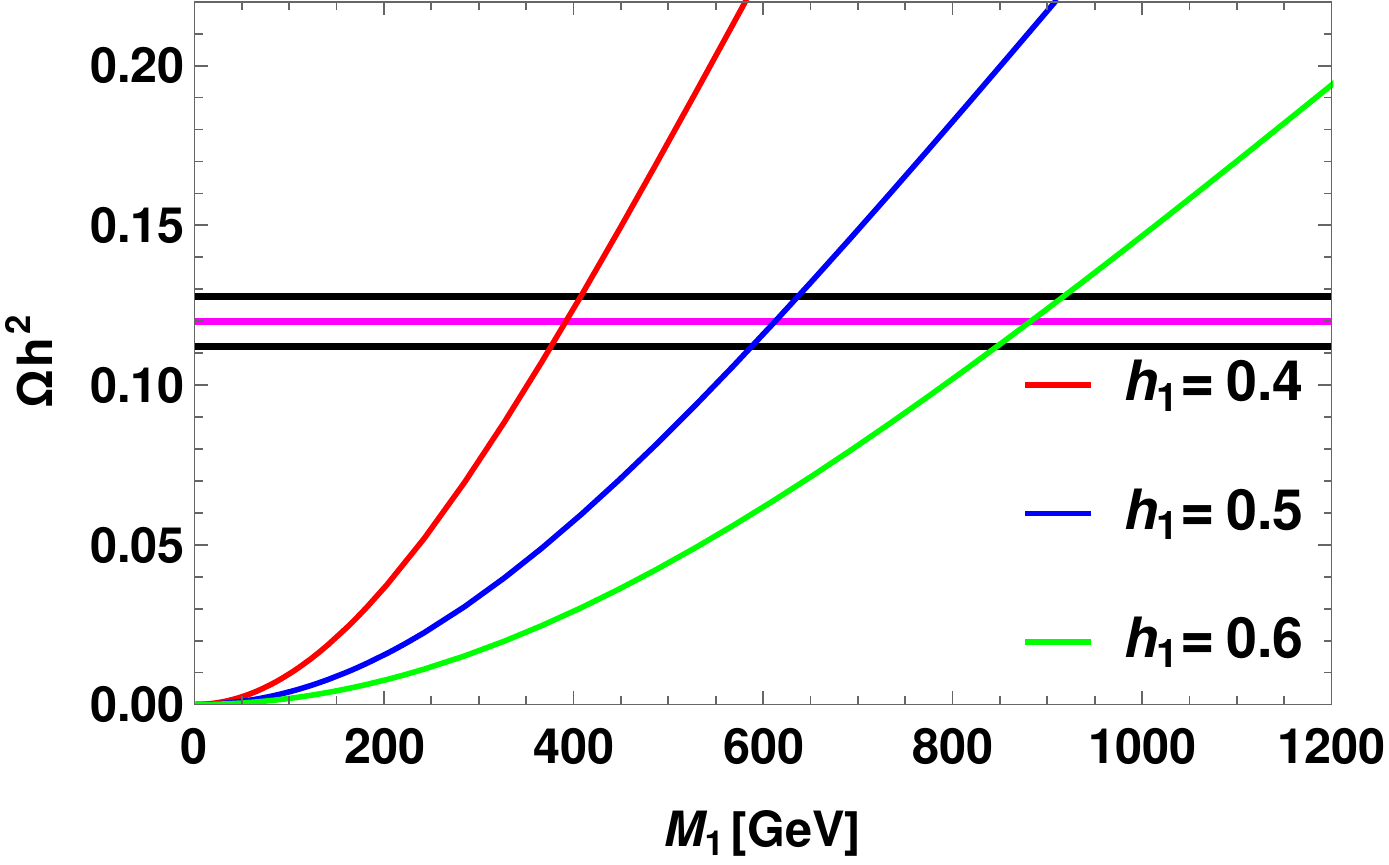}
\caption{Variation of relic abundance with DM mass for various values of $h_1$ at $r_1 = 0.5$ (left panel) and $r_1 = 0.6$ (right panel) where the horizontal line (Magenta) represents the central value of the relic density whereas the black lines denote their corresponding 3$\sigma$ range. }
\end{center}
\label{omegavsM}
\end{figure}
%%%%%%%%%%%%%%%%%%%%%%%%%%%%%%%%%%%%%%%%%%%%%%%%%%%%%%%%%%%%%%%%%%%%%%%%%%%%%

As the light neutrinos oscillate in flavor, one loop diagrams contribute to lepton flavor
violating decays. We now further constrain the parameter space of the model using these decays.
\section{Lepton flavour violating decays}
The observation of neutrino oscillations has provided unambiguous signal for lepton number violation in the neutral lepton sector, even though the individual
lepton number is conserved in the SM of the electroweak interaction. The evidence of light neutrino masses and mixing and the violation of family lepton number could in principle allow flavor changing neutral current (FCNC) transitions in the charged lepton sector as well, such as $\ell_\alpha\to\ell_\beta\gamma$ and $\ell_{\alpha}\to
\ell_{\beta}\overline{\ell_{\beta}}\ell_{\beta}$.

\begin{figure}[t]
\begin{center}
\includegraphics[scale=0.45]{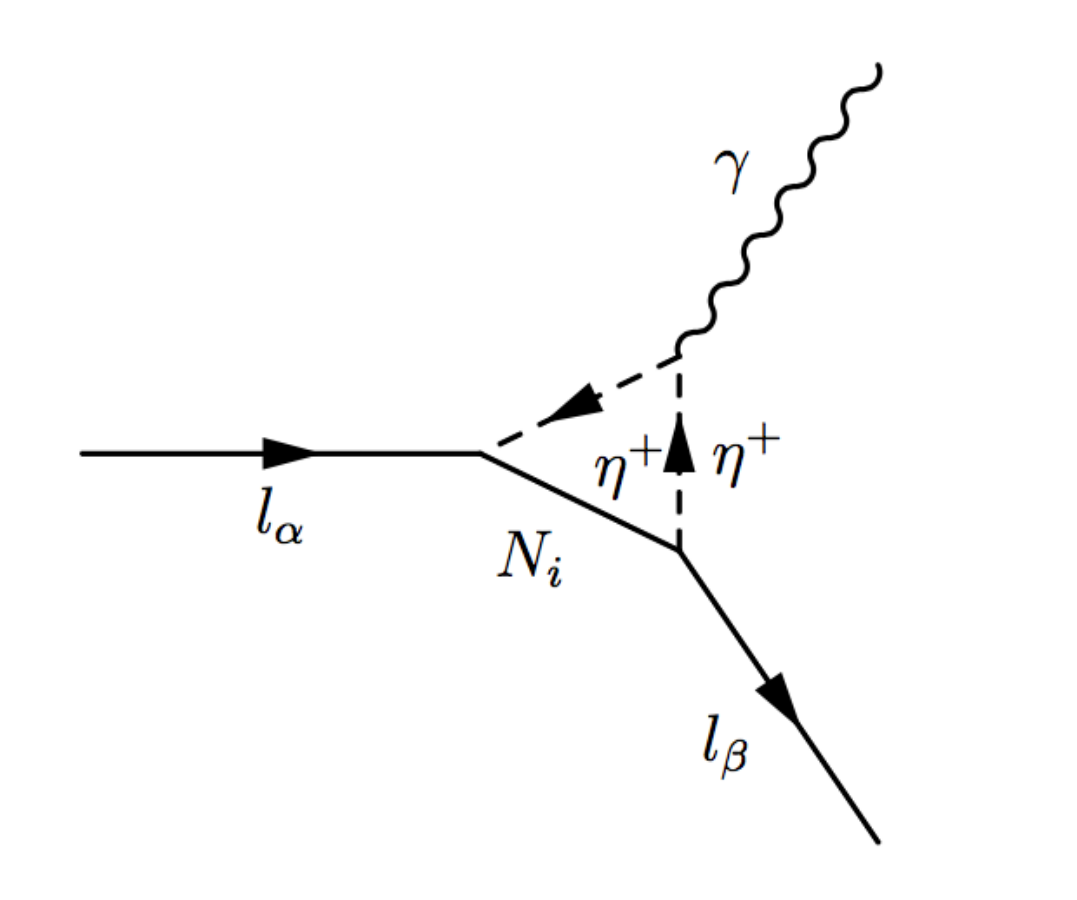}
\quad
\includegraphics[scale=0.45]{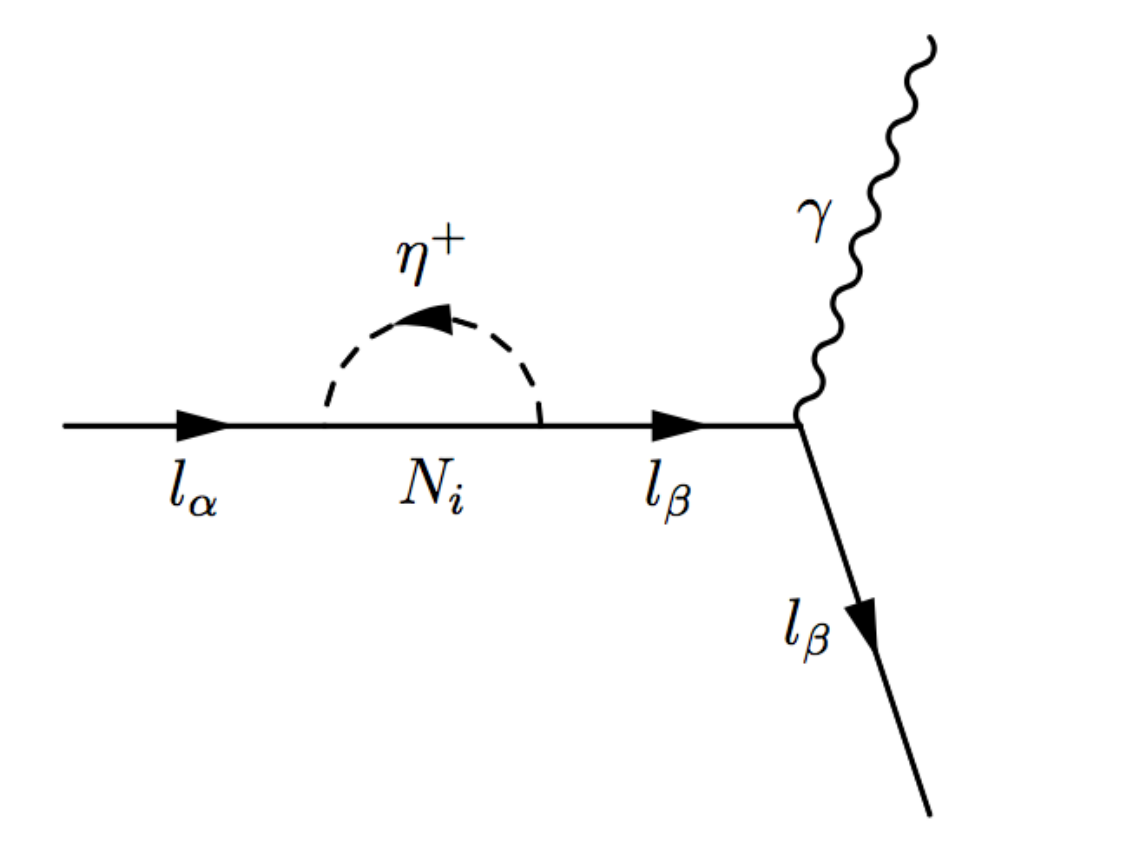}
\quad
\includegraphics[scale=0.45]{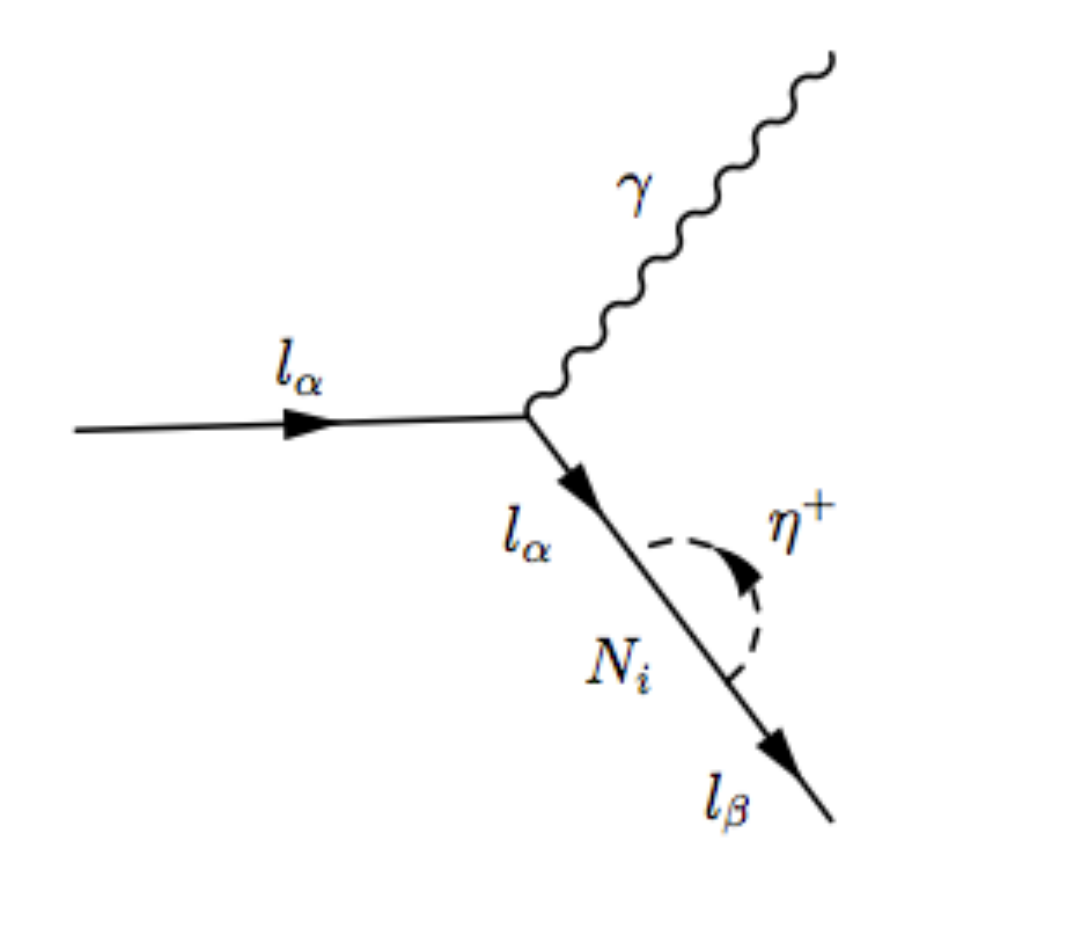}
\caption{Diagrams contributing to $l_{\alpha} \to l_{\beta} \gamma$.}
\label{mueg}
\end{center}
\end{figure}

The expression for the branching ratio of lepton flavor
violating decay process $\ell_\alpha\to\ell_\beta\gamma$ written in terms of dipole form factor $A_D$ is given by \cite{vicente1}
\begin{equation}
\mathrm{Br}(\ell_\alpha\to\ell_\beta\gamma)=
\frac{3(4\pi)^3\alpha_{\mathrm{em}}}{4
G_F^2}|A_D|^2 \mathrm{Br}
\left(\ell_\alpha\to\ell_\beta\nu_\alpha
\overline{\nu_\beta}\right),
\label{mu2egamma}
\end{equation}
where $\alpha_{\mathrm{em}}=e^2/4\pi$ is the electromagnetic fine
structure constant, $G_F$ is the Fermi constant and $\alpha (\beta)$ represents the lepton flavor.
The diagrams contributing to $A_D$ are shown in Fig. \ref{mueg} and the expression is given by
%%%%%%%%%%%%%%%%%%%%%%%%%%%%%%%%%%%%%%%%%%%%%%%%%%%%%%%%%%
%%%%%%%%%%%%%%%%%%%%%%%%%%%%%%%%%%%%%%%%%%%%%%%%%%%%%%%%%
\begin{equation}
A_D = \sum_{i=1}^3\frac{h_{i\beta}^*h_{i\alpha}}
{2(4\pi)^2}\frac{1}{m_{0}^2}
F_2\left(r_i\right).
\label{ad}
\end{equation}
Here expression for the loop function $F_2(x)$ is given in appendix A and for simplicity we consider $\lambda_4 \ll \lambda_3$, thus we get $\eta^+$ and $\eta^0$ to be  degenerate \cite{suematsu}. Applying the flavour structure (\ref{flav}), the relation  (\ref{mu2egamma}) becomes
\begin{equation}
\mathrm{Br}(\mu \to e\gamma)=
\frac{3\alpha_{\mathrm{em}}}{64\pi
G_F^2m_0^4}\left|\left(h_{2}^2 - 
0.68h_{1}^2 \right)F_2\left(r_1\right) + (3.56 h_{3}^2)F_2\left(r_3\right) \right|^2.
\end{equation}
We consider $r_3 > 1$, $M_1 < 2 {~\rm TeV}$ and $M_3, m_0 < 8 {~\rm TeV}$ and thus we work in the mass regime $M_1 \simeq M_2 < m_0 < M_3$.  Of all the lepton flavor violating (LFV) decays, the decay channel $\mu \to e\gamma$ provides most stringent constraint on the parameter space of this model. 
%%%%%%%%%%%%%%%%%%%%%%%%%%%%%%%%%%%%%%%%%%%%%%%%%%%%%%%%%%%%%%%%%%%%%%%%%%%%
\begin{figure}[htb]
\includegraphics[width=7 cm ,height=5.0 cm, clip]{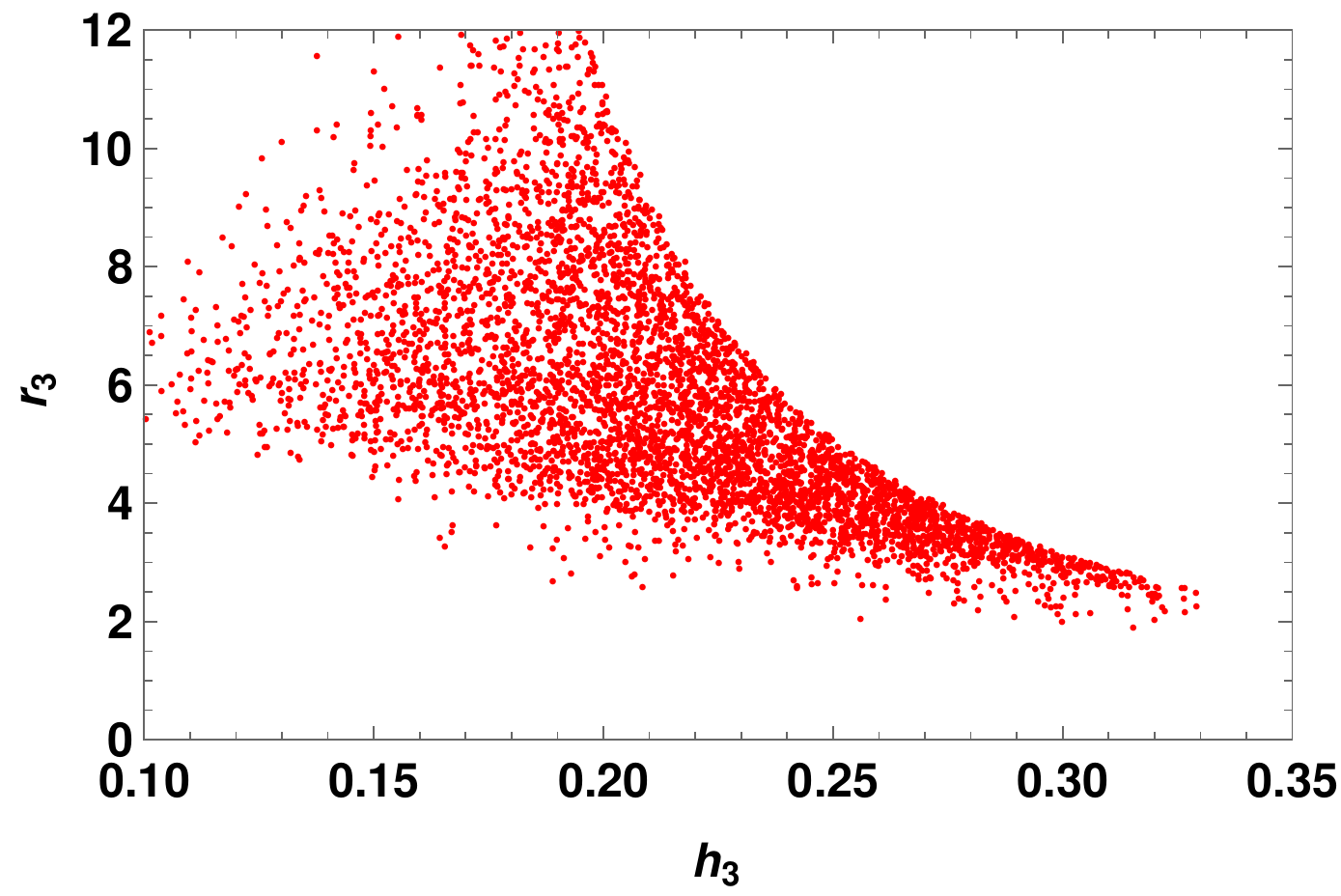}
\quad
\includegraphics[width=7 cm ,height=5.0 cm, clip]{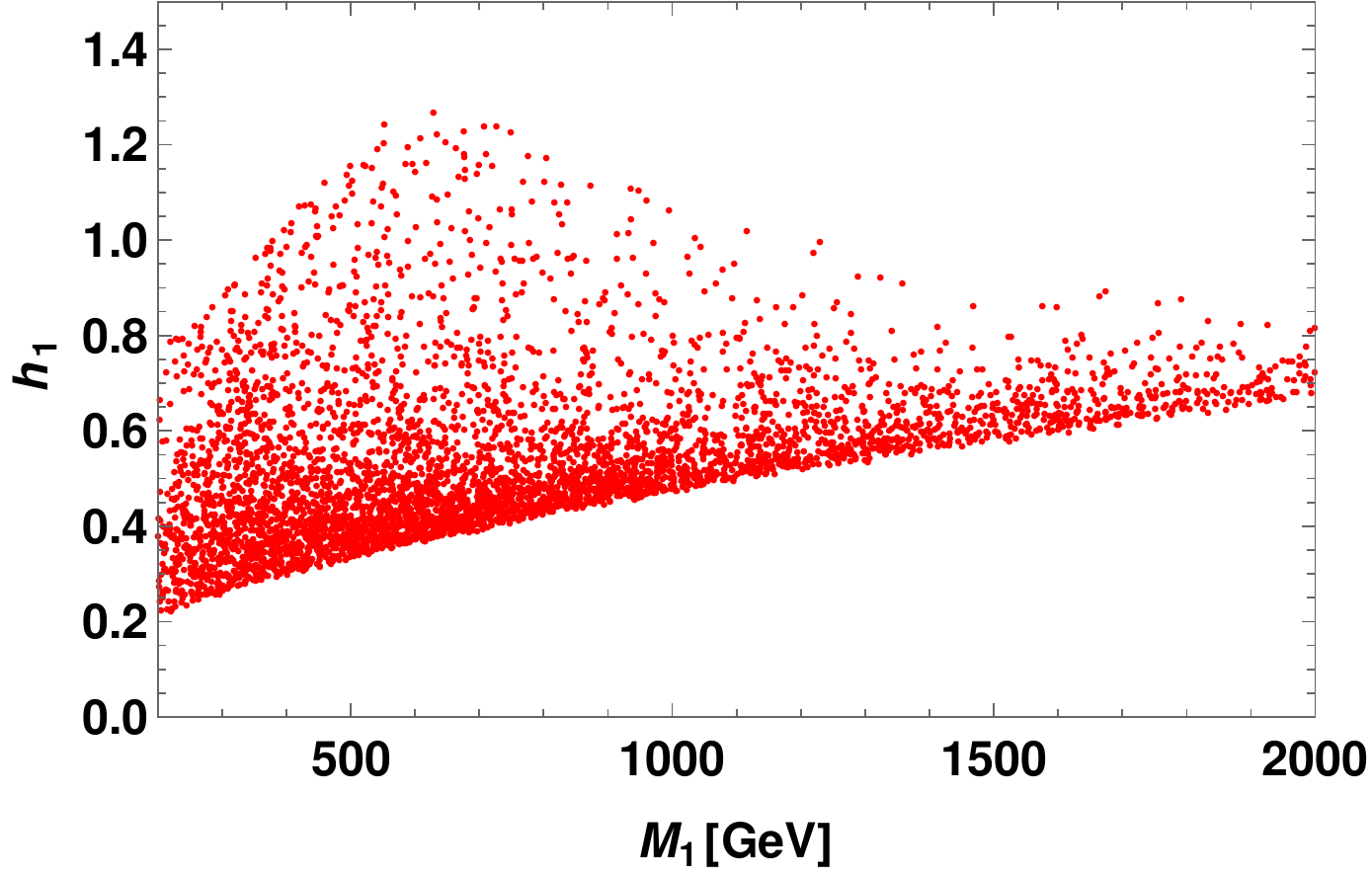}
\caption{Parameter space of $h_3$ and $r_3$ (left panel) and variation of $h_1$ with $M_1$ (right panel) consistent with neutrino oscillation data, relic density and Br$(\mu \to e\gamma)$.}
\label{h1h3}
\end{figure}
%%%%%%%%%%%%%%%%%%%%%%%%%%%%%%%%%%%%%%%%%%%%%%%%%%%%%%%%%%%%%%%%%%

Imposing the constraints from neutrino mass square differences, relic abundance and current upper bound on ${\rm Br}(\mu \to e\gamma)$ \cite{pdg}, Fig. \ref{h1h3} (left panel) shows the allowed region in ($h_3$, $r_3$) parameter space of the model. From the figure, one can conclude that the lower bound on $r_3$ is 2 (i.e., $r_3 > 2$) and the upper bound on $h_3$ is 0.33 (i.e., $h_3 < 0.33$ ). Fig. \ref{h1h3} (right panel) depicts the variation of $h_1$ with the mass of DM. It shows that Br$(\mu \to e\gamma)$ excludes the values above $1.2$ for  $h_1$. Now taking all the constraints from the flavor and dark sector, one can tabulate the allowed parameter space shown in Table. II.
%%%%%%%%%%%%%%%%%%%%%%%%%%%%%%%%%%%%%%%%%%%%%%%%%%%%%%%
\begin{table}[htb]
\begin{center}
\vspace*{0.1 true in}
\begin{tabular}{|c|c|c|}
\hline
 Parameters & Range  \\
\hline
$r_1$  &~ $0.2 \to 1$ ~\\
$r_3$  &~ $2 \to 12$ ~\\
$|h_1|$ &~ $0.2 \to 1.2 $ ~\\
$|h_2|$ &~ $0.2 \to 1.0 $ ~\\
$|h_3|$ &~ $0.1 \to 0.33 $ ~\\
\hline
\end{tabular}
\end{center}
\caption{Scotogenic model parameters with their range.}
\end{table}
%%%%%%%%%%%%%%%%%%%%%%%%%%%%%%%%%%%%%%%%%%%%%%%%%%%%%%%

\begin{figure}[htb]
\includegraphics[width= 7.0 cm ,height=5.0 cm, clip]{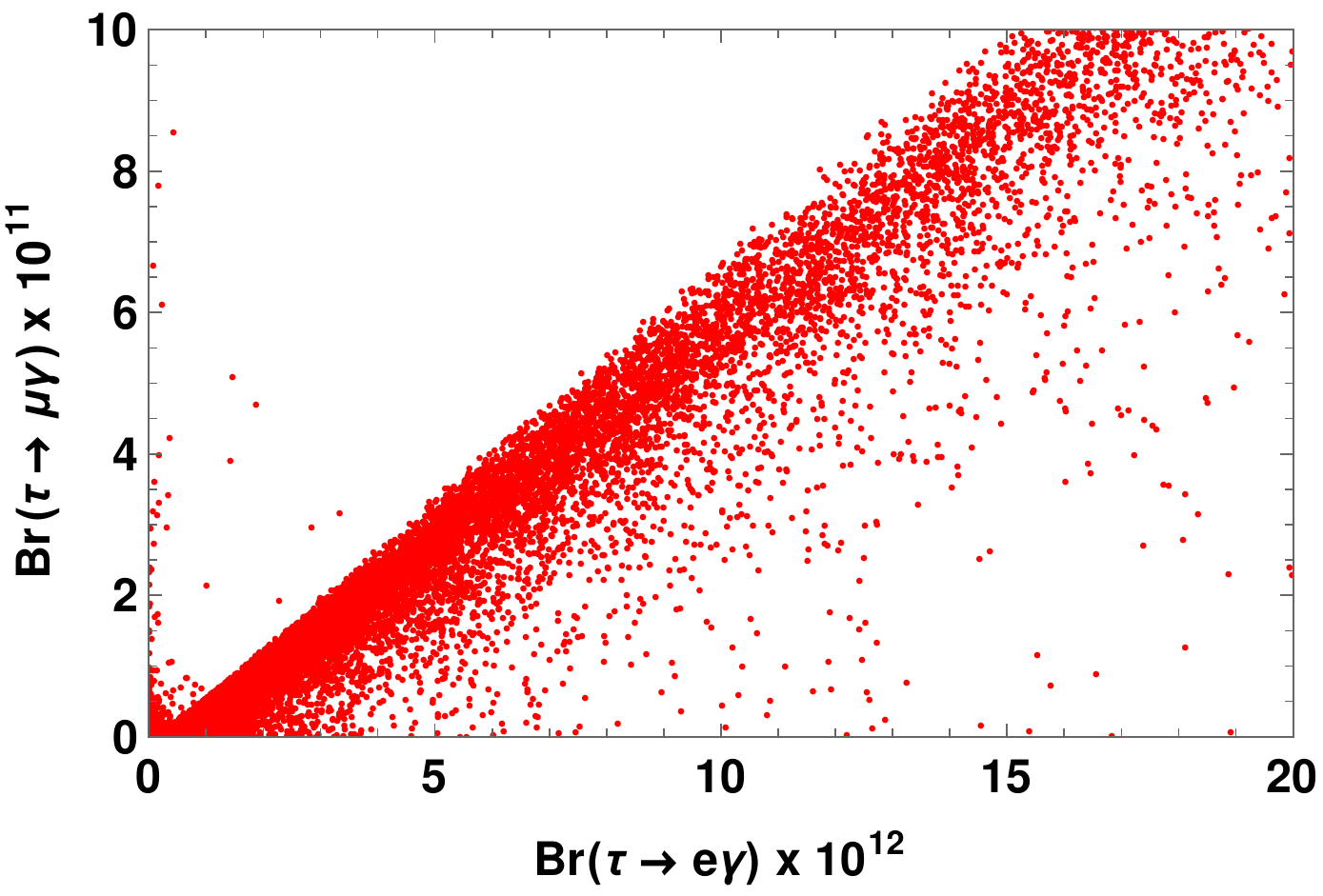}
\caption{Correlation plot between $\rm{Br}(\tau \to e\gamma)$ and $\rm{Br}(\tau \to \mu \gamma)$}
\label{tauemug}
\end{figure}

We follow the similar procedure to compute the branching ratios of $\tau \to e\gamma$ and $\tau \to \mu \gamma$ decays. Using the allowed parameter space given in Table-II, we show in Fig. \ref{tauemug} the correlation plot between $\rm{Br}(\tau \to e\gamma)$ and $\rm{Br}(\tau \to \mu \gamma)$. In our analysis, we have used the  measured branching ratios for $\mu \to \nu_\mu e \bar{\nu}_e$, $\tau^- \to \nu_\tau \mu^- \bar{\nu}_\mu$ and
$ \tau^- \to \nu_\tau e^- \bar{\nu}_e $  processes from \cite{pdg} as
\bea
{\rm Br}( \mu \to \nu_\mu e \bar{\nu}_e)&=& 100 \% \;,\nn\\
{\rm Br}( \tau \to \nu_\tau \mu \bar{\nu}_\mu)&=& (17.41 \pm 0.04) \% \;,\nn\\
{\rm Br}( \tau \to \nu_\tau e \bar{\nu}_e)&=& (17.83 \pm 0.04) \% \;.
\eea
%%%%%%%%%%%%%%%%%%%%%%%%%%%%%%%%%%%%%%%%%%%%%%%%%%%%%%%%%%%%%%%%%%%%%%%%%%%%
\\
%%%%%%%%%%%%%%%%%%%%%%%%%%%%%%%%%%%%%%%%%%%%%%%%%%%%%%%%%%%%%%%%%
Now we study lepton flavor violation in 3-body decays. As discussed in Ref.\cite{vicente1}, these decays get contributions from three types of loop diagrams namely : $\gamma$-penguin, $Z$-penguin and box diagrams. The branching ratio for
$\ell_\alpha \to 3 \, \ell_\beta$ in scotogenic model is given by \cite{vicente1}
\begin{eqnarray}
\text{Br}\left(\ell_{\alpha}\to
\ell_{\beta}\overline{\ell_{\beta}}\ell_{\beta}\right)&=&
\frac{3(4\pi)^2\alpha_{\mathrm{em}}^2}{8G_F^2}
\left[|A_{ND}|^2
  +|A_D|^2\left(\frac{16}{3}\log\left(\frac{m_\alpha}{m_\beta}\right)
  -\frac{22}{3}\right)+\frac{1}{6}|B|^2\right.\nonumber\\
 &&\left.
    +\left(-2 A_{ND} A_D^{*}+\frac{1}{3}A_{ND} B^*
  -\frac{2}{3}A_D B^*+\mathrm{h.c.}\right)\right]\nonumber\\
&&\times \, \mathrm{Br}\left(\ell_{\alpha}\to\ell_{\beta}\nu_{\alpha}
\overline{\nu_{\beta}}\right) \, .
\end{eqnarray}
%%%%%%%%%%%%%%%%%%%%%%%%%%%%%%%%%%%%%%%%%%%%%%%%%%%%%%%%%
\begin{figure}[t]
\begin{center}
\includegraphics[scale=0.7]{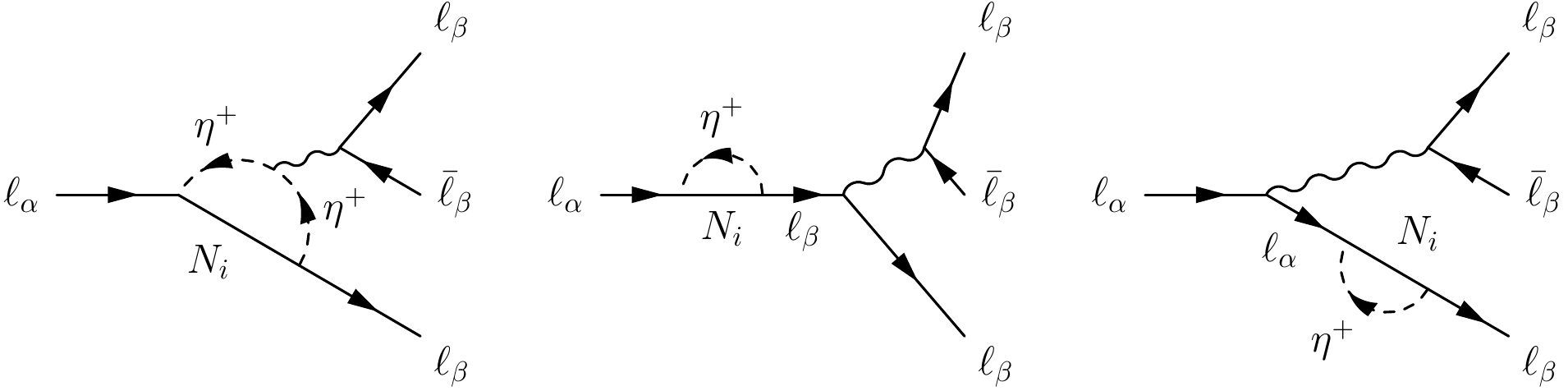}
\caption{Penguin diagram contributions to
  $\ell_\alpha \to 3 \, \ell_\beta$ with the mediator (wavy line) denotes either a photon or a Z-boson.}
\label{peng}
\end{center}
\end{figure}
%%%%%%%%%%%%%%%%%%%%%%%%%%%%%%%%%%%%%%%%%%%%%%%%%%%%%%%%%%%
The coefficient $A_D$ denotes photon dipole contributions  given in eqn. (\ref{ad}), whereas the coefficient  ${A_{ND}}$ represents the form factor with the photonic non-dipole contributions given by
\begin{equation}
A_{ND}=\sum_{i=1}^3\frac{h_{i\beta}^*h_{i\alpha}}
{6(4\pi)^2}\frac{1}{m_{0}^2}
G_2\left(r_i\right).
\label{and}
\end{equation}
Here $G_2(x)$ is a loop function is given in appendix A.
$Z$-pengiun diagrams shown in Fig. \ref{peng} give negligible contribution to the decay width as explained in \cite{vicente1, vicente2}.
Apart from photon dipole and non-dipole penguin contributions, the box diagrams shown in Fig. \ref{box} also contribute to the decay width given by
\begin{eqnarray}
 B = \frac{1}{(4\pi)^2{e^2}m_{0}^2} 
 \sum_{i,\:j=1}^3\left[ \frac{1}{2} D_1(r_i, r_j) h_{j \beta}^* h_{j \beta}
	   h_{i \beta}^* h_{i \alpha} + r_i r_j
	   D_2(r_i, r_j)) h_{j \beta}^* h_{j \beta}^* h_{i \beta}
	   h_{i \alpha}  \right].
\end{eqnarray}
The loop functions $D_1(x,y)$ and $D_2(x,y)$ are provided in appendix A.
%%%%%%%%%%%%%%%%%%%%%%%%%%%%%%%%%%%%%%%%%%%%%%%%%%%%%%%%%%%%%%%%
\begin{figure}[t]
\begin{center}
\includegraphics[scale=0.5]{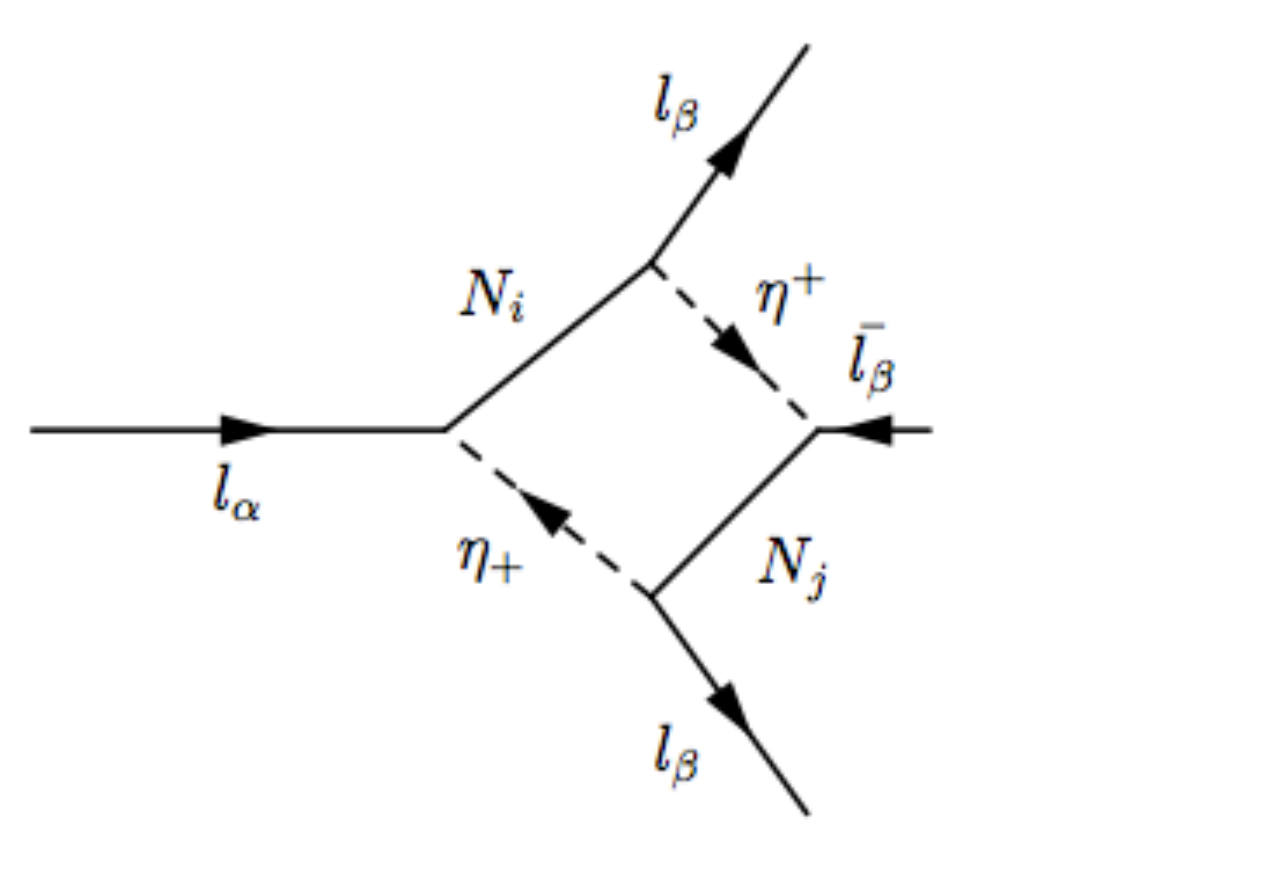}
\quad
\includegraphics[scale=0.5]{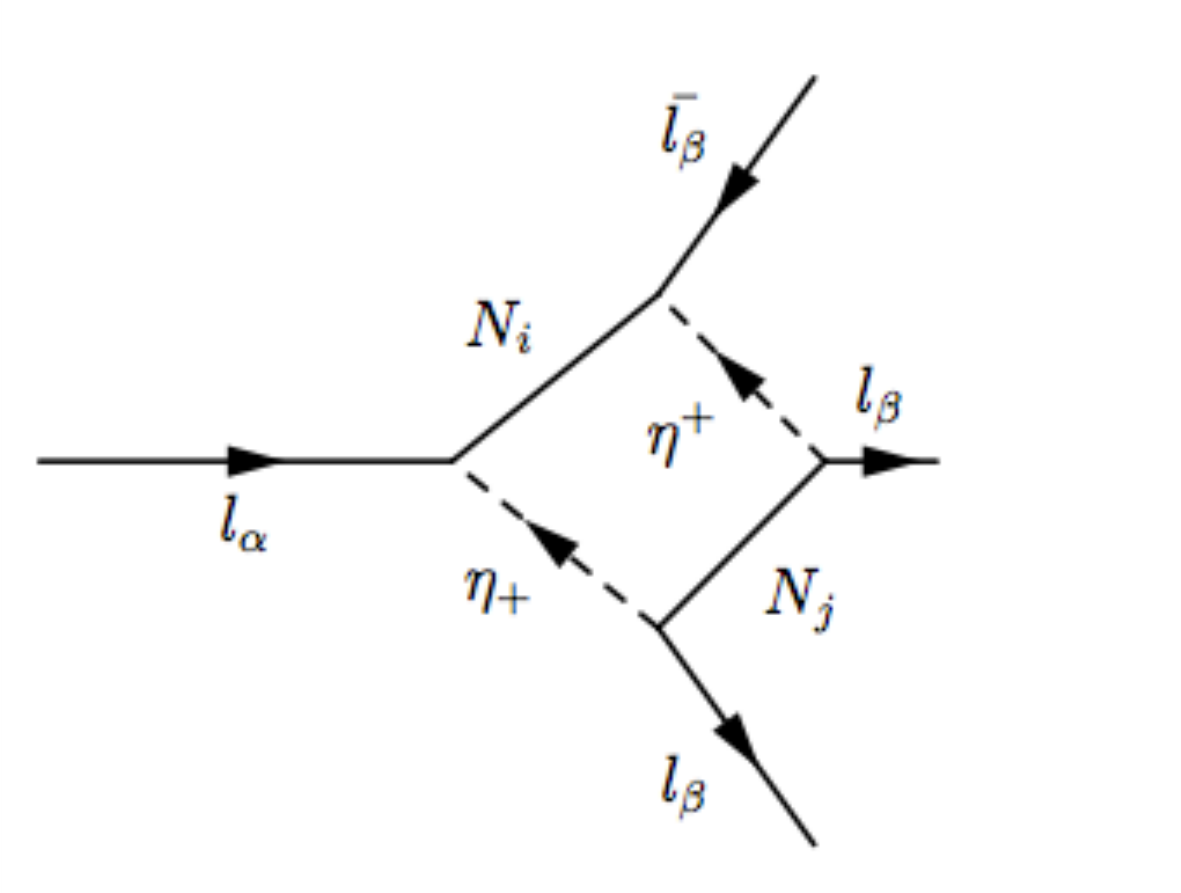}
\caption{Box diagram contributions to
  $\ell_\alpha \to 3 \, \ell_\beta$.}
\label{box}
\end{center}
\end{figure}
%%%%%%%%%%%%%%%%%%%%%%%%%%%%%%%%%%%%%%%%%%%%%%%%%%%%%%%%%%%%%

\begin{figure}[htb]
\includegraphics[width=7 cm ,height=5.0 cm, clip]{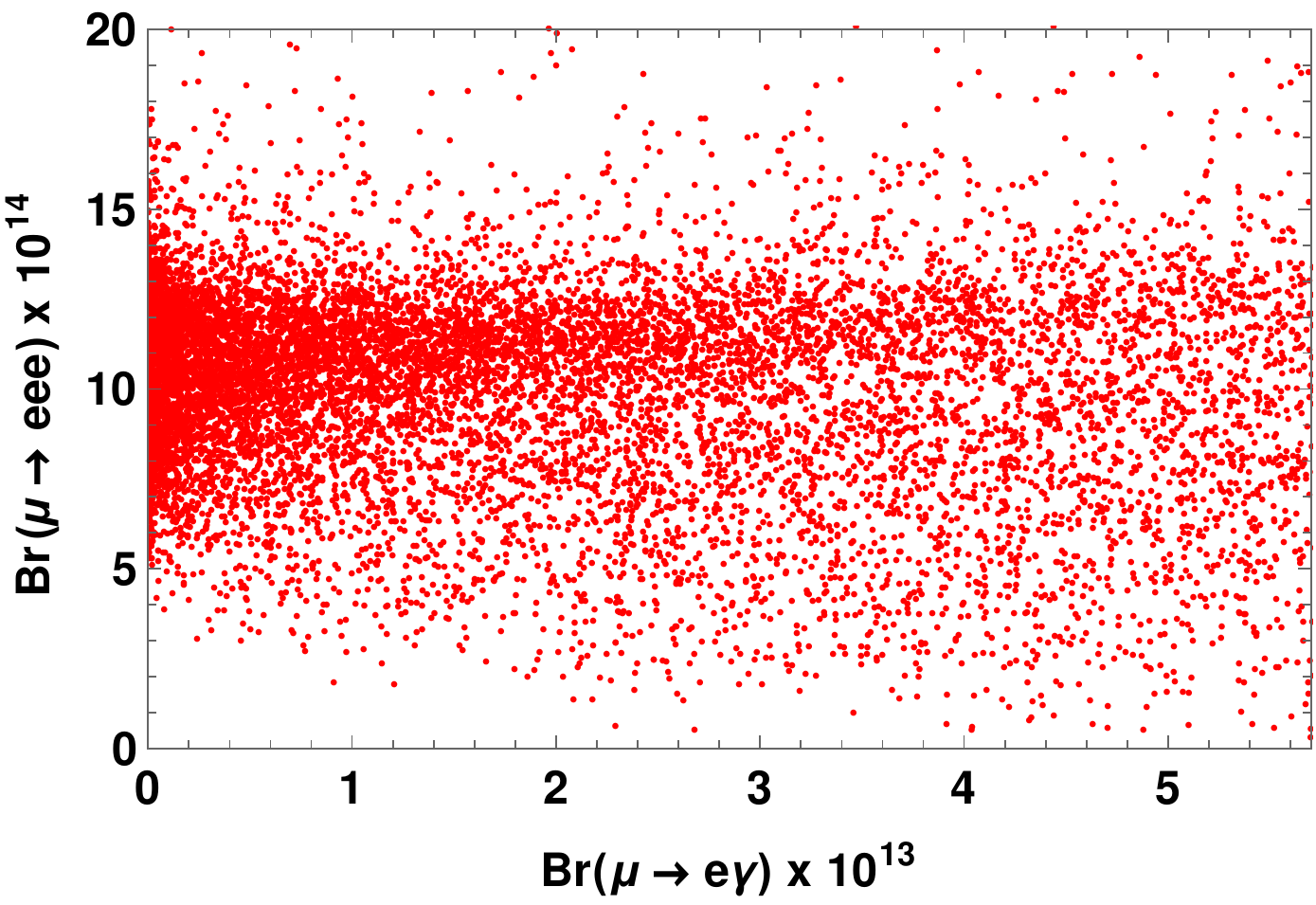}
\quad
\includegraphics[width=7 cm ,height=5.0 cm, clip]
{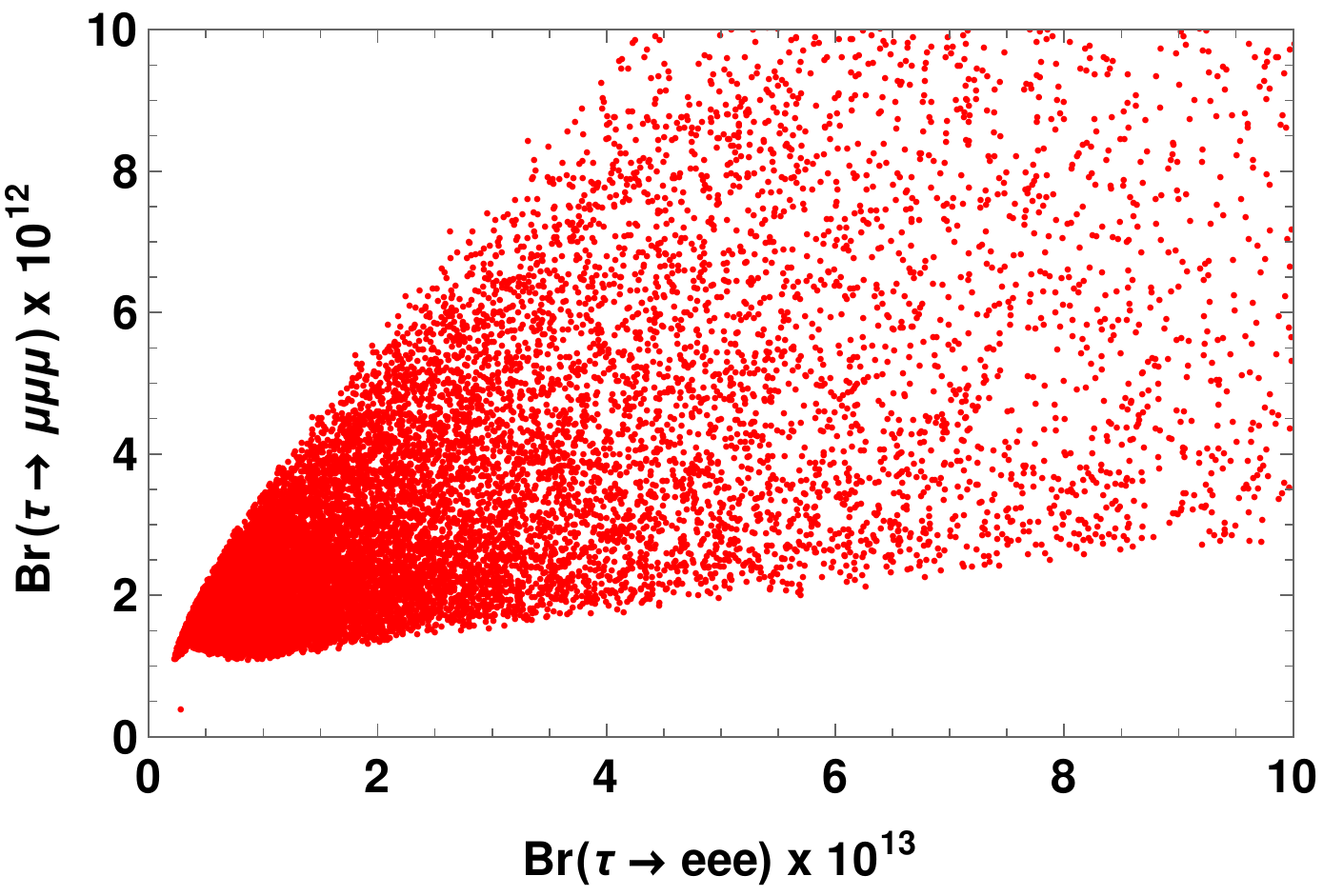}
\caption{Correlation plot between ${\rm Br}(\mu \to eee)$ and ${\rm Br}(\mu \to e\gamma)$ (left panel) and between ${\rm Br}(\tau \to \mu\mu\mu)$ and ${\rm Br}(\tau \to eee)$ (right panel)}
\label{mu}
\end{figure}

Using the allowed parameter space from Table- II, we show in Fig. \ref{mu} the correlation plot between $\mu \to e\gamma$ and $\mu \to eee$ (left panel). Similarly the right panel in Fig. \ref{mu} depicts the correlation plot between branching ratios of $\tau \to eee$ and $\tau \to \mu \mu \mu$. From these figures we conclude that all the obtained branching ratios in the viable parameter space are within the experimental limits.
%%%%%%%%%%%%%%%%%%%%%%%%%%%%%%%%%%%%%%%%%%%%%%%%%%%%%%%%%%%%%%%%%%%%%%%%%%%%
%%%%%%%%%%%%%%%%%%%%%%%%%%%%%%%%%%%%%%%%%%%%%%%%%%%%%%%%%%
\section{Summary and Conclusion}
To summarize, in this paper we have considered scotogenic model which is an extension of standard model with an additional inert scalar doublet and three heavy Majorana right-handed neutrinos. It is a noval scenario connecting neutrino physics and dark matter. We have diagonalized the neutrino radiative mass matrix using the TBM matrix with an additional perturbed matrix as a rotation matrix in 13 plane. The mixing angles are chosen ($\theta=35^\circ$ and $\varphi=12^\circ$) to accommodate sizeable $\theta_{13}$.  Working in a degenerate heavy neutrino mass spectrum, we have obtained the flavor structure favourable to explain normal neutrino mass ordering. Choosing the lightest among the odd particles as dark matter, we have computed the relic abundance including the coannihilation effects. Scanning over the entire parameter space and applying the constraints from neutrino oscillation data, dark matter observables and bounds from lepton flavor violating decays such as $\ell_\alpha\to\ell_\beta\gamma$ and $\ell_\alpha \to 3 \, \ell_\beta$, we have shown the suitable range for various parameters in the model.
 
%%%%%%%%%%%%%%%%%%%%%%%%%%%%%%%%%%%

\appendix
\section{Loop functions}
The loop functions used in LFV decays are given by
\begin{eqnarray}
F_2(x) &=& \frac{1-6x^2+3x^4+2x^6-6x^4 \log x^2}{6(1-x^2)^4}, \\
G_2(x) &=& \frac{2-9x^2+18x^4-11x^6+6x^6 \log x^2}{6(1-x^2)^4}, \\
D_1(x,y) &=& - \frac{1}{(1-x^2)(1-y^2)} - \frac{x^4 \log x^2}{(1-x^2)^2(x^2-y^2)} -
 \frac{y^4 \log y^2}{(1-y^2)^2(y^2-x^2)}, \\
D_2(x,y) &=& - \frac{1}{(1-x^2)(1-y^2)} - \frac{x^2 \log x^2}{(1-x^2)^2(x^2-y^2)} -
 \frac{y^2 \log y^2}{(1-y^2)^2(y^2-x^2)}.
\end{eqnarray}
In the limit $y\to x$, the functions $D_1$ and $D_2$ become 
\begin{eqnarray}
&&D_1(x,x)=\frac{-1+x^4-2x^2\log{x^2}}{(1-x^2)^3},\\
&&D_2(x,x)=\frac{-2+2x^2-(1+x^2)\log{x^2}}{(1-x^2)^3}.
\end{eqnarray}

%%%%%%%%%%%%%%%%%%%%%%%%%%%%%%%%%%% 


\begin{thebibliography}{99}

\bibitem{t1}
P.~Minkowski,
{Phys. Lett. B}, {\bf 67}: 421 (1977);
R. N.~Mohapatra and G.~Senjanovic,
{Phys. Rev. Lett.} {\bf 44}: 912 (1980);
M.~Gell-Mann, P.~Ramond, and R.~Slansky (1980), print-80-0576 (CERN);
J.~Schechter and J. W. F.~Valle,
Phys. Rev. D,  {\bf 22}: 2227 (1980).

\bibitem{t2} R. N. Mohapatra and G. Senjanovic, Phys. Rev. D, {\bf 23}: 165 (1981); G. Lazarides, Q. Shafi and C. Wetterich, Nucl. Phys. B, {\bf  181}: 287 (1981); C. Wetterich, Nucl. Phys. B, {\bf 187}: 343 (1981); B. Brahmachari and R. N. Mohapatra, Phys. Rev. D, {\bf 58}: 015001 (1998);
S. Antusch and S. F. King, Phys. Lett. B, {\bf 597}: 199 (2004);
R. N. Mohapatra, Nucl. Phys. Proc. suppl. {\bf 138}: 257 (2005).


\bibitem{t3} R. Foot, H. Lew, X. G. He et al, Z. Phys. C, {\bf 44}: 441 (1989).

\bibitem{ma} E.~Ma, Phys. Rev. D, {\bf 73}: 077301 (2006), arXiv:hep-ph/0601225.

\bibitem{pmns} B. Pontecorvo, Sov. Phys. JETP, {\bf 7}: 172 (1958);
Z. Maki, M. Nakagawa and S. Sakata, Prog. Theor. Phys, {\bf 28}: 870 (1962).

\bibitem{daya-bay1}  F.~P.~An {\it et al} (DAYA-BAY Collaboration),  Phys. Rev. Lett. {\bf 108}: 171803 (2012), arXiv:1203.1669.

\bibitem{daya-bay2}  F.~P.~An {\it et al} (DAYA-BAY Collaboration), Chin. Phys. C, {\bf 37}: 011001 (2013), arXiv:1210.6327.

\bibitem{reno}  J.~K.~Ahn {\it et al} (RENO Collaboration), {  Phys.  Rev.  Lett. } {\bf 108}: 191802 (2012), 
arXiv:1204.0626.

\bibitem{t2k}   K.~Abe {\it et al} (T2K Collaboration), Phys. Rev. D, {\bf 88}:  032002 (2013), arXiv:1304.0841.

\bibitem{osc} D.~Forero, M.~Tortela and J.~Valle, Phys. Rev. D, {\bf 90}: 093006 (2014), arXiv:1405.7540.

\bibitem{planck}	P. A. R. Ade et al (Planck Collaboration), Astron. Astrophys. {\bf 571}: A1 (2014), arXiv:1303.5062.

\bibitem{lopez} L.~Lepoz~Honorez, E.Nezri, J.~F.~Oliver et al, JCAP, {\bf 02}: 28 (2007), arXiv:hep-ph/0612275.

\bibitem{barbieri} R.~Barbieri, L.~E.~Hall and V.~S.~Rychkov, Phys. Rev. D, {\bf  74}: 015007 (2006), arXiv:hep-ph/0603188.


\bibitem{Gustafsson} M. Gustafsson, PoS CHARGED, {\bf 2010}: 030 (2010), arXiv:1106.1719.

\bibitem{suematsu} Daijiro Suematsu, Takashi Toma, Tetsuro Yoshida, Phys. Rev. D, {\bf 79}: 093004 (2009), arXiv:0903.0287.

\bibitem{schmidt} D. Schmidt, T Schwetz, T. Toma, Phys. Rev. D, {\bf 85}: 073009 (2012), arXiv:1201.0906.

\bibitem{vicente1} T. Toma and A. Vicente, JHEP, {\bf 01}: 160 (2014), arXiv:1312.2840.

\bibitem{vicente2} A. Vicente, C E. Yaguna, JHEP, {\bf 02}: 144 (2015), arXiv:1412.2545.

%\bibitem{raidal} E.~Ma and M.~Raidal, Phys. Rev. Lett. {\bf 87} (2001) 011802.

\bibitem{sym}G. Altarelli, F. Ferugilo, Rev. Mod. Phys. {\bf 82}: 2701 (2010); G. Altarelli, F. Ferugilo, L. Merlo,
E. Stamou, JHEP, {\bf 08}: 021 (2012), arXiv:1205.4670; S. F. King and C. Luhn, Rept. Prog.
Phys. {\bf 76}: 056201 (2013), arXiv:1301.1340; S. F. King, A. Merle, S. Morisi et al, New Journ. Phys. {\bf 16}: 045018 (2014); H. Isimori et al., Prog. Theor. Phys.
Suppl. {\bf 183}: 1 (2010).

\bibitem{tbm}P. F. Harrison, D. H. Perkins, and W.G. Scott, Phys. Lett. B, {\bf 458}: 79 (1999); Phys. Lett.
B, {\bf 530}: 167 (2002); Z. Z. Xing, Phys. Lett. B, {\bf 533}: 85 (2002); P. F. Harrison and W. G. Scott,
Phys. Lett. B, {\bf 535}: 163 (2002); Phys. Lett. B, {\bf 557}: 76 (2003); X.-G. He and A. Zee, Phys.
Lett. B, {\bf 560}: 87 (2003);  L. Wolfenstein, Phys. Rev. D, {\bf 18}: 958 (1978); Y. Yamanaka,
H. Sugawara, and S. Pakvasa, Phys. Rev. D, {\bf 25}: 1895 (1982); D, {\bf 29}: 2135(E) (1984); N. Li and
B.-Q. Ma, Phys. Rev. D, {\bf 71}: 017302 (2005), arXiv:hep-ph/0412126.

\bibitem{sruthilaya} 	
M. Sruthilaya, C. Soumya, K.N. Deepthi et al, New J. Phys. {\bf 17}:   083028 (2015), arXiv:1408.4392.

\bibitem{griest} K.~Griest and D.~Seckel, Phys. Rev. D, {\bf 43}: 3191 (1991).

\bibitem{pdg} K. A. Olive  {\it et al}, Particle Data Group Collaboration, Chin. Phys. C, {\bf 38}: 090001 (2014).

\end{thebibliography}
\end{document}